\def\be{\begin{equation}}
\def\ee{\end{equation}}

\documentclass[12pt,a4paper,epsfig]{iopart}
\usepackage{epsfig}
\usepackage{graphicx}
\usepackage[below]{placeins} 
\usepackage{textcomp}
\usepackage{color}
\usepackage{amssymb}

\makeatletter
\renewcommand{\@fnsymbol}[1]{%
   \ifcase#1\or\textasteriskcentered\or\textsection\or\textdagger
   \or\textdaggerdbl\or\textparagraph\or\textbardbl
   \or\textasteriskcentered\textasteriskcentered
   \or\textdagger\textdagger\or\textdaggerdbl\textdaggerdbl
   \else\@ctrerr\fi}
\makeatother

\begin{document}
\title[Effect of the attachment of contacts]
      {Effect of the attachment of ferromagnetic contacts on the conductivity and giant magnetoresistance of graphene nanoribbons}
\author{S. Krompiewski}%
\address{Institute of Molecular Physics, Polish Academy of
Sciences, ul.~M.~Smoluchowskiego 17, 60179 Pozna{\'n}, Poland
}%
\date{\today}


 \begin{abstract}

Carbon-based nanostructures and graphene, in particular, evoke a
lot of interest as new promising materials for nanoelectronics and
spintronics. One of the most important issue in this context is
the impact of external electrodes on electronic properties of
graphene nanoribbons (GNR). The present theoretical method is
based on the tight-binding model and a modified recursive
procedure for Green's functions. The results show that within the
ballistic transport regime, the so called end-contacted geometry
(of minimal GNR/electrode interface area), is usually more
advantageous for practical applications than its side-contacted
counterpart (with a larger coverage area), as far as the
electrical conductivity is concerned. As regards the giant
magnetoresistance coefficient, however, the situation is exactly
opposite, since spin-splitting effects are more pronounced in the
lower conductive side-contacted setups.
\end{abstract}
\maketitle

\section{Introduction}

Different contact configurations affect the conductivity, and
thereby also giant magnetoresistance (GMR) in the case of
ferromagnetic electrodes. Here we study two most common
geometries, \emph{viz.} the end-contacted (EC) geometry and the
side-contacted (SC) one. In the former the interface area is
minimal, whereas in the latter a finite coverage area exists.
Theoretically, it is easier to work with EC setups rather than,
experimentally more relevant, SC ones. The former approach might
be justified when a nanostructure/electrode coupling is strong
(good transparency of interfaces, weak chemical bonding). There
are experimental evidences supporting this view. In particular,
Mann et al. \cite{Mann_nlet2003} were able to control a length of
CNT (carbon nanotube) under a metal contact and showed that
ballistic transport in Pd/CNT occurred basically at the edges.
Moreover, in similar systems, no observable difference in I-V
characteristics between side-contacts and end-contacts was found
\cite{Song_nanotech2009}. In general, however the contact
configuration does matter; as shown in \cite{Nemec_2008} for Pd
and Ti contacted carbon nanostructures. Likewise, \emph{ab initio}
studies in \cite{Matsuda_JPC2010} show that in the case of
ultra-small graphene nanoribbons, the EC configuration outperforms
the SC one, since a large interface area may worsen the contact
conductance.

As concerns ferromagnetic electrodes, it has been shown that in
order for a transition metal/graphene junction to have good spin
transport properties it is necessary that no strong chemical bonds
(engaging graphene's $\pi$ electrons) are formed
\cite{Giovannetti_prl2008, Maassen_2011}. Although in the case of
typical ferromagnetic electrodes this requirement is hardly
fulfilled \cite{Nagashio_Apl10}, adsorption of graphene (e.g. on
Ni or Co) can be effectively weakened by introducing a protective
interfacial monolayer  of Cu \cite{Giovannetti_prl2008}.
Noteworthy, high-performance
ferromagnetic-metal/graphitic-nanostructure contacts have already
been reported for Ni$_{1-x}$Pd$_x$ with $x \sim 0.3 \div 0.5$ (see
\cite{Cottet_sst06} and references therein).


Although the EC geometry is much more frequently used by
theoreticians just for reasons of computational convenience
(\cite{Blanter_prb2007}-\cite{Krompiewski_prb2009}), the issue of
side-contacted nanostructures has already also been raised, see
\cite{Nemec_2008,Matsuda_JPC2010,Barraza-Lopez10,Xia_nnan11}
(paramagnetic electrodes) and
\cite{Maassen_2011,Nagashio_Apl10,Stokbro} (ferromagnetic
electrodes).
 In this study we treat both the geometries on the equal footing,
and show how the corresponding transport characteristics
(conductivity, shot noise Fano factor and giant magnetoresistance)
compare with each other.

   \begin{center}
  \begin{figure}[t]
\includegraphics [width=6in] {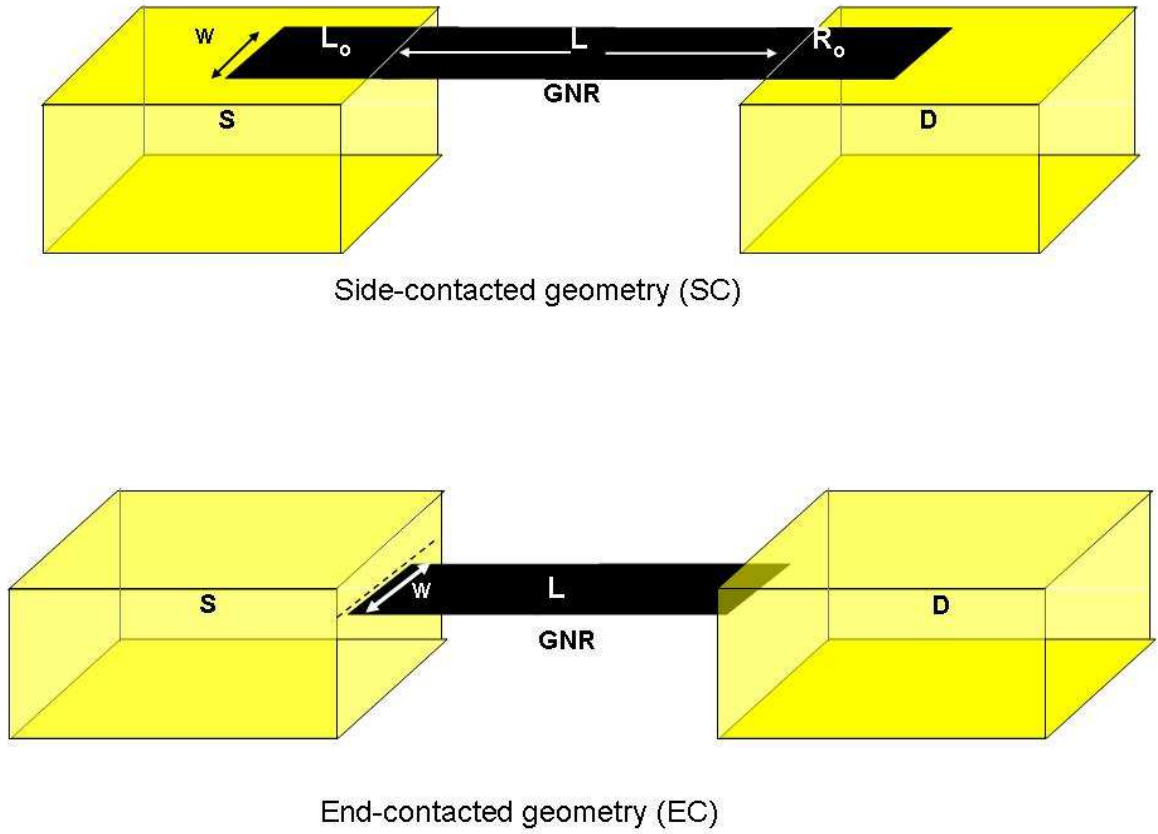}
\caption{\small (Color online) Schematics of the side-contacted
and end-contacted setups. S and D denote the source and drain
electrodes. The GNR is drawn in black, and W, L, $L_0$ and $R_0$
stand for its width, suspended length and electrode-supported
lengths, respectively.}
       \label{fig1}
   \end{figure}
   \end{center}

\newpage

\section{Method and Modelling}

We study graphene nanoribbons with infinite 3-dimensional metallic
electrodes. Details on modelling and the computation method may be
found in \cite{Krompiewski_prb2009,Krompiewski_Nano2011}. However,
for completeness, the most important definitions are listed below.

\begin{eqnarray*} \label{GMR}
 g_{ss',s''}  &=& (L/W)\frac{e^2}{h} Tr(T_{ss',s''}), \\
 F_{ss'} & = &  1- \frac{Tr(T^2_{ss',\uparrow} +T^2_{ss',\downarrow})}{Tr(T_{ss',\uparrow} +T_{ss',\downarrow})} , \\
GMR&=&100 \left(1- \frac{g_{\uparrow
\downarrow,\uparrow}+g_{\uparrow
\downarrow,\downarrow}}{g_{\uparrow \uparrow,\uparrow}+g_{\uparrow
\uparrow,\downarrow}}\right).
\end{eqnarray*}

      \begin{center}
  \begin{figure}[b]
\includegraphics [width=6in] {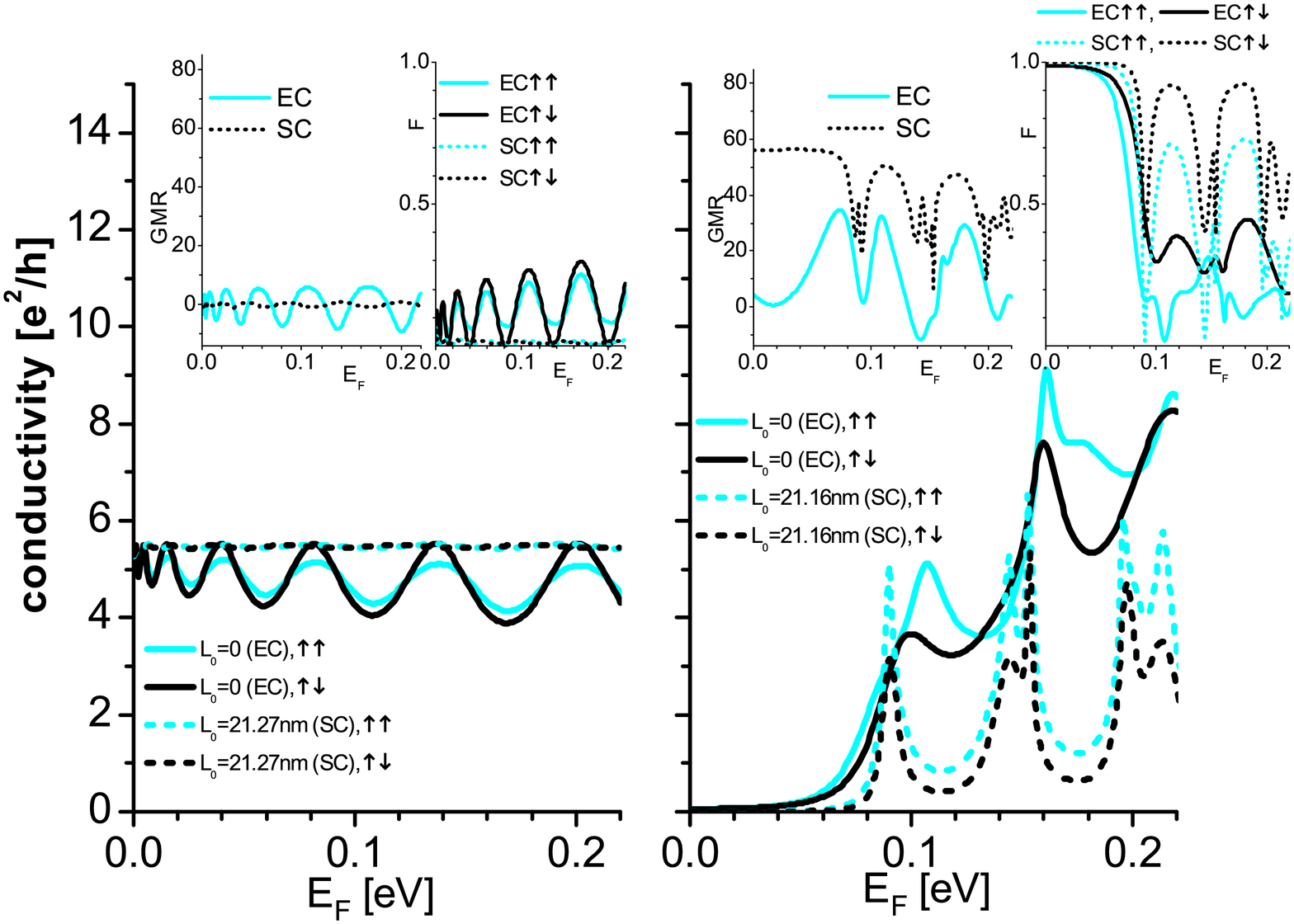}
\caption{\small (Color online) The metallic zigzag GNRs and the
semiconducting armchair GNRs of comparable sizes (W $\sim 8$nm, L
$\sim$ 22nm, and A$\sim$1/3), with the indicated
electrode-supported lengths $L_0$: conductivities (main panels),
GMR coefficients (left Insets), and shot noise Fano factors (right
Insets). The parallel (antiparallel) electrode alignment is marked
with $\uparrow \uparrow$ ($\uparrow \downarrow$).}
       \label{fig2}
   \end{figure}
   \end{center}

Here $g$ is the conductivity of the GNR, having width $W$ and
length $L$, for the parallel or antiparallel alignments of
electrode magnetizations ($ss'=\uparrow \uparrow$, $\uparrow
\downarrow$, respectively). $T_{ss',s''}$, in turn, stands for the
transmission matrix of $s''$-spin electrons in the $ss'$
configuration. Finally, GMR is a giant magnetoresistance
coefficient.

The modified recursion procedure for SC setups is carried out
stepwise. The vertical contribution to the transmission matrix is
related with the self-energy $\Sigma_S^I=T_{I0,IS} \; g_S^I \;
T_{IS,I0}$, where I=L (R) is the left (right) electrode index, IS
(I0) enumerates electrode (graphene) interface atoms and $g_S$ is
the surface Green's function matrix of the electrodes. The
starting local Green's functions are defined as
$g_{0}^L=(E-D_{L0}-\Sigma_S^L)^{-1}$,
$g_{N+1}^R=(E-D_{R0}-\Sigma_S^R)^{-1}$, where L0 (R0) are the
electrode-supported parts of the GNR (with the matrices of hoping
integrals D), whose suspended part extends from the periodicity
unit number 1 up to N.  Figure 1 schematically shows the systems
under consideration, and defines the end-contacted and
side-contacted geometries.

The calculations have been carried out using a single-band ($\pi$
orbital) tight-binding Hamiltonian, with the GNRs strongly coupled
to infinite ferromagnetic external electrodes (assumed to be 50\%
spin-polarized), and the adopted approach is based on the Landauer
type formalism. As mentioned in Sec.1 quasi-ballistic transport in
carbon nanostructures has been reported in numerous papers (e.g.
\cite{Xia_nnan11},\cite{Liang_Nat01}-\cite{Du_nnan08}). Here we
ensure a good transparency of the GNR/contact interface by
defining the hopping integral across the interface as a
geometrical mean of the respective hopping integrals of the GNR
and the electrode (\emph{cf.}
\cite{Blanter_prb2007,Krompiewski_prb2009}).

 \begin{center}
     \begin{figure}[b]
\includegraphics [width=6in] {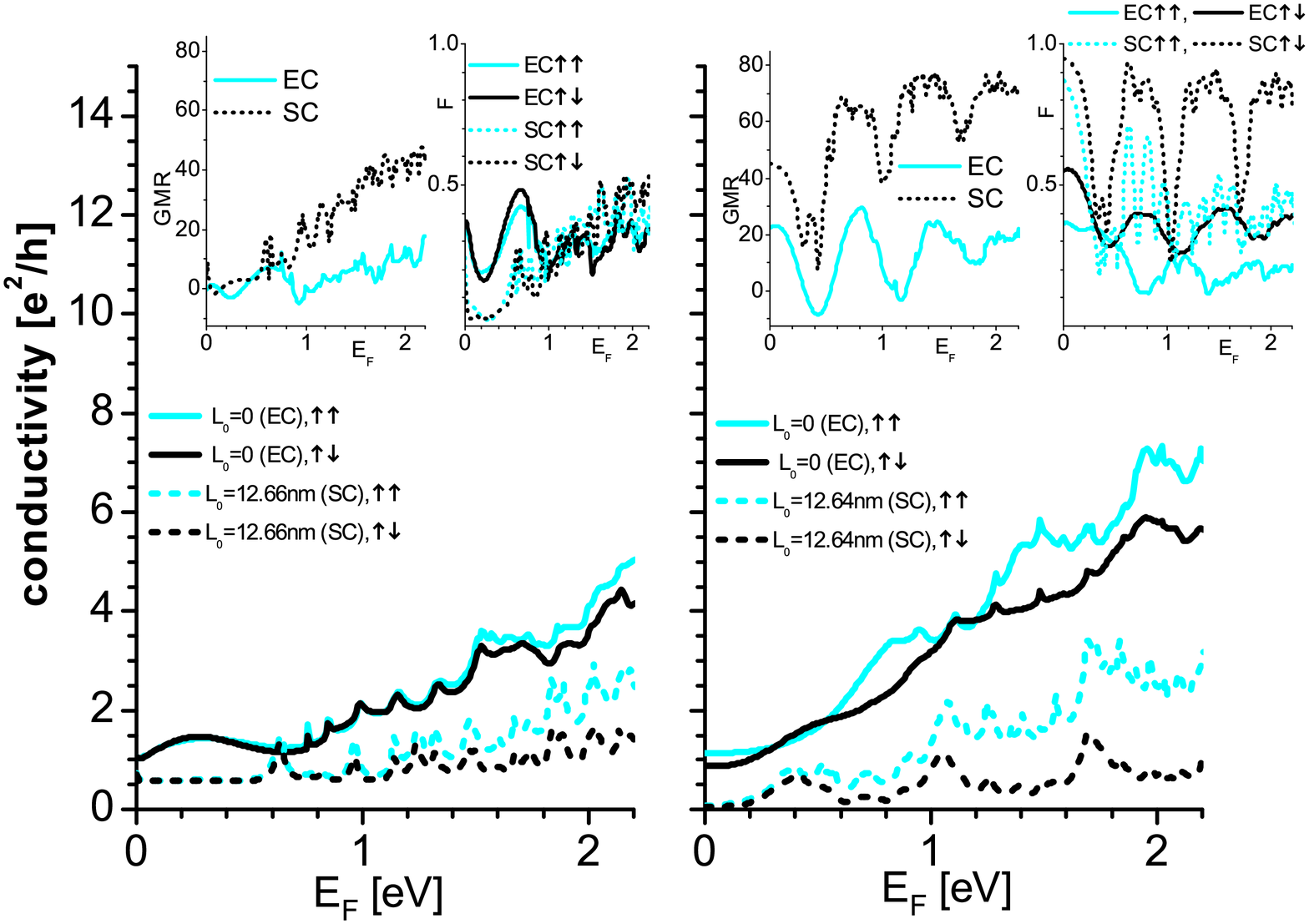}
\caption{\small (Color online) As Fig.~\ref{fig2} but for the
short systems with the aspect ratio A$\sim$3.5.}
       \label{fig3}
   \end{figure}
   \end{center}

     \begin{figure}[t]
     \begin{center}
\includegraphics [width=6in] {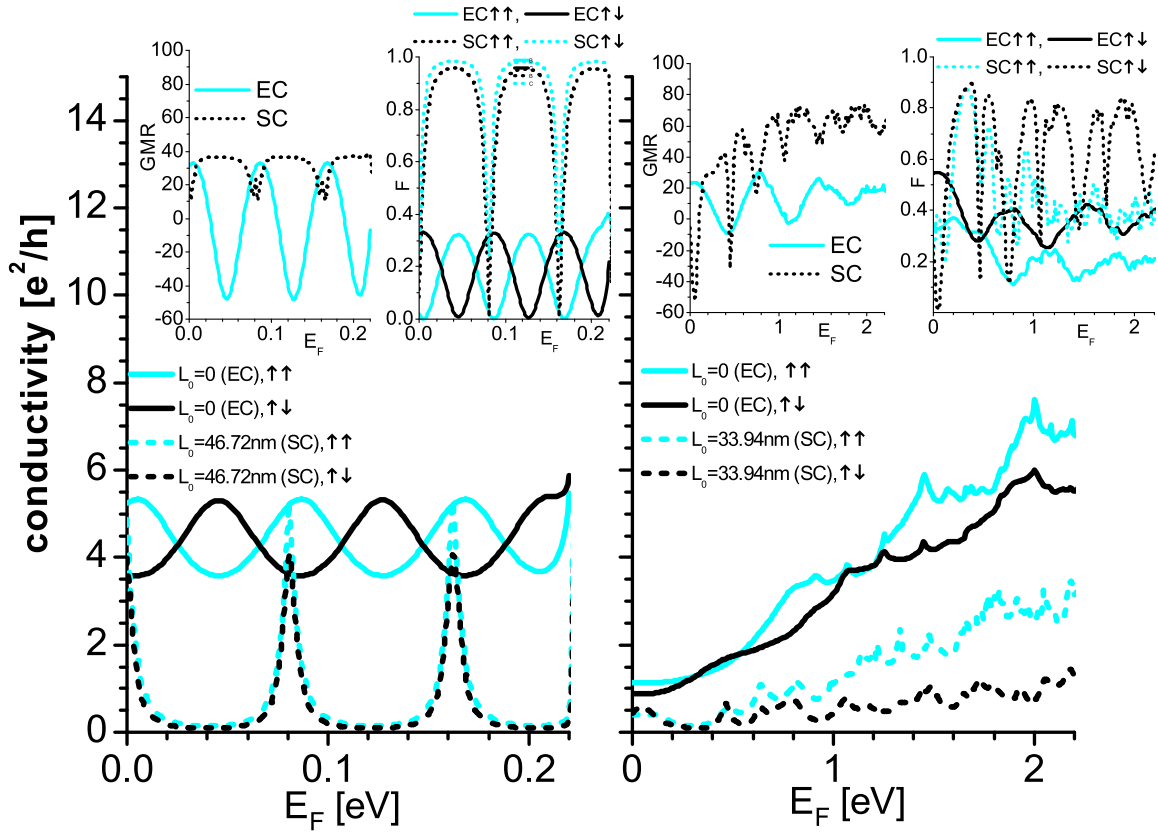}
\caption{\small (Color online) Spin-dependent conductivities, GMRs
and Fano factors for metallic armchair GNRs in the end- and side-
contacted geometry.}
       \label{fig4}
     \end{center}
   \end{figure}

\section{Discussion of the Results}

In the SC case the transport properties depend on the interface
area. For small area interfaces the results are similar to those
corresponding to the EC geometry, otherwise the results converge
with the increasing area. In what follows, the SC results, which
will be presented, correspond to the GNR/electrode coverages large
enough to guarantee the convergence. The representative results
are presented in Figs.~2-4, as a function of the GNR Fermi energy
which is experimentally controllable by gate voltage. The aim is
to elucidate the role of the interface character (EC \emph{vs.}
SC), the aspect ratio (A=W/L) and the current flow direction
(zigzag \emph{vs.} armchair) on spin-dependent transport in
2-terminal setups with ferromagnetic contacts. The convention
adopted here is the following: solid lines apply to the EC
geometry, dashed lines - to the SC geometry, whereas the cyan
(brighter) curves and the black ones correspond to the parallel
and antiparallel alignments of the ferromagnetic electrodes,
respectively. The figures 2-3 correspond to the metallic zigzag
GNRs and the semiconducting armchair GNRs of comparable sizes. In
shorthand notation the systems in question are: zz19-90 and
ac35-52 (Fig.~2), and zz19-10 and ac35-6 (Fig.~3), where the first
number denotes the GNR's width and the second one (after the
hyphen) refers to the suspended length, all these numbers are
expressed in periodicity units appropriate for the given
direction. Remarkably, for a sufficiently long coverage length
($L_0$) the long zigzag GNR reveals a saturation effect as
concerns the spin-dependent conductivity (at conductance$=A \cdot$
conductivity $\sim 2e^2/h$), resulting in gradually vanishing of
both GMR and F. These features may be understood, as due to
development of a low-energy perfectly conducting channel with no
backscattering counterpart (\emph{cf.} a paramagnetic electrode
case of \cite{Krompiewski_Nano2011}). Interestingly the results
for the EC and SC configuration do not differ much in the zigzag
case; in other cases the respective differences are larger, unless
the electrode/GNR overlap is substantially smaller than the
indicated SC values of $L_0$ in the figures. It has been recently
shown by \emph{ab initio} calculations that Ni/GNR/Ni setups, with
very small overlap lengths (up to $8 \AA$), have almost identical
contact resistivities which moreover are independent of the
current orientation \cite{Stokbro}. It should be stressed in this
context that the present phase-coherent transport theory
overestimates interference processes, which in reality may be
suppressed by diffusive and dephasing mechanisms. Anyway, within
this approach the conductivity does depend on the current
propagation direction unlike \cite{Stokbro} but in accord with
\cite{Zhou_jpcm10,Zhou_jpcm11}. Most probably this discrepancy
comes from different aspect ratios of the studied systems.
Noteworthy, the present theory correctly reproduces the so-called
universal minimum conductivity value predicted theoretically
\cite{Tworzydlo} and confirmed experimentally
\cite{Miao_Sci07,Danneau08}, since the low-energy conductivities
corresponding to the high aspect ratio EC systems are close to $4
e^2/(h \pi$) (see Figs. 3-4).

Figure~4 shows the results for nominally metallic armchair GNR
ac34-52 and ac34-6 of different length. The former (longer)
reveals pronounced Fabry-P{\'e}rot oscillations, resulting from
the linear energy spectrum in this case. Both the systems behave
similarly to the other armchair systems discussed so far, in
particular the conductivities corresponding to the ES geometry are
clearly higher than those referring to the SC geometry. At the
same time, the EC Fano factors are, on average, lower than the SC
ones. So, the EC geometry seems to be optimal for nanoelectronic
applications. However the accompanying spin effects - spin
splitting of the conductivities and the GMR - are most often the
strongest in the side contacted systems.

The difference in conductances between the EC and SC
configurations provides information about a contact resistance. If
both the conductances are roughly equal to each other than the
contact is characterized just by the channel width, otherwise the
areal component comes into play \cite{Nagashio_Apl10}. It turns
out that the contact resistances may be vanishingly small,
especially at gate-controllable resonances (see Figs 2 and 4).
Incidentally, very small contact resistances have been
experimentally found, as well, in nearly ballistic graphene/metal
junctions \cite{Xia_nnan11}. In the case of more complicated
electrodes than those studied here, extra contributions to the
contact resistance can be important (e.g. spreading resistance
\cite{Garcia_prb08,Venugopa_apl10} ). It is so, when an STM (STS)
tip is used as an electrode and/or there is a constriction in the
system. In the context of spintronics related studies of graphene,
it should be also mentioned that apart from the GMR also the
spin-torque-transfer (STT) problem is crucial for magnetic memory
applications. These two akin phenomena are due to a long
spin-relaxation length in carbon nanostructures and, in principle,
the present method has a potential to describe the STT, too;
obviously when one of the electrodes is a non-massive free layer
(see \cite{Zhou_jpcm10,Zhou_jpcm11,Yokoyama_prb11} for SST in
EC-type setups).

\section{Summary}

In conclusion, the present quasi-ballistic theory shows that,
spin-dependent transport through GNRs, depends on their aspect
ratio, the GNR/electrode interface, and the current direction. It
turns out that the conductivity of the semiconducting GNRs, as
well as that of the nominally metallic armchair GNRs, is clearly
higher in the EC geometry than in the SC geometry. However in the
case of long zigzag GNRs, which are always metallic, both the
geometries yield comparable conductivities. As concerns the GMR
coefficients and the Fano factors, they are always anticorrelated
with the conductivities, and typically the SC GMR effect is higher
than the EC one. Moreover, practically in all the cases the range
of the appearance of the so-called inverse (negative) GMR is
strongly reduced in the SC geometry.

The new insight provided by this study is that high aspect ratio
(wide and short) GNRs appear good candidates for interconnects in
miniaturized electronic devices due to their preferential EC
(edge) type conductance. As concerns miniaturization in
spintronics, the short EC armchair GNRs, with GMR$\sim 20\%$ seem
also quite attractive.

\vspace{0.5 cm}

 \noindent \textbf{Acknowledgments}

This work was supported by the Polish Ministry of Science and
Higher Education as a research project No. N N202 199239 for
2010-2013.

\end{document}